\documentclass[twocolumn,floatfix,times,reprint,twocolappendix]{aastex701}

\usepackage{graphicx}

\usepackage{subcaption}
\usepackage{xcolor}
\usepackage{amsmath}
\usepackage{amsfonts}
\usepackage{amssymb}
\usepackage{latexsym}

\usepackage{dcolumn}
\usepackage{bm} 
\usepackage{float} 
\usepackage{booktabs}
\usepackage{enumitem} 
\usepackage{natbib}

\usepackage{cleveref}

\usepackage[english]{babel}
\usepackage[expansion=all,babel]{microtype}
\usepackage{mathtools}

\usepackage{multirow}
\usepackage[flushmargin]{footmisc} 

\usepackage{newtxtext,newtxmath}

\definecolor{MB}{RGB}{255,99,71} 

\newcommand{\Msun}{\mathrm{M}_\odot}

\begin{document}

\title{Rotochemical heating of neutron stars in 2D}

\author[orcid=0000-0002-7326-7270]{Mikhail V. Beznogov}
\affiliation{National Institute for Physics and Nuclear Engineering (IFIN-HH), RO-077125 Bucharest, Romania}
\email{mikhail.beznogov@nipne.ro}

\author[0000-0002-6029-4712]{J\'er\^ome Novak}
\affiliation{Universit\'e de Strasbourg, CNRS, Observatoire astronomique de Strasbourg, UMR 7550, F-67000 Strasbourg, France}
\email{jerome.novak@obspm.fr}

\author[orcid=0000-0003-2498-4326]{Dany Page}
\affiliation{Instituto de Astronomía, Universidad Nacional Autónoma de México, Ciudad de México, CDMX 04510, Mexico}
\email{page@astro.unam.mx}

\author[orcid=0000-0001-8421-2040]{Adriana R. Raduta}
\affiliation{National Institute for Physics and Nuclear Engineering (IFIN-HH), RO-077125 Bucharest, Romania}
\email{araduta@nipne.ro}

\correspondingauthor{Mikhail V. Beznogov}
\email{mikhail.beznogov@nipne.ro}

\begin{abstract}
The matter inside a spinning-down neutron star (NS) undergoes a slow compression that may drive it out of chemical equilibrium and, consequently, trigger non-equilibrium nuclear reactions that can heat up the star.
This phenomenon is called rotochemical heating, and it is relatively well-investigated in the literature.
Historically, it has been studied almost exclusively via 1D spherically symmetric models.
However, rapidly rotating NSs are not spherically symmetric.
We present pioneering simulations of rotochemical heating in full General Relativity in 2D axially symmetric models without relying on the slow rotation approximation.
We explicitly follow thermal, mechanical, and chemical evolution of rotating NSs as they spin down taking into account both radial and angular distributions of temperature, composition, and mechanical structure.
The simplest model is employed: only nucleonic composition, only the non-equilibrium modified Urca process as the main heating/cooling mechanism, and no superfluidity. 
We analyze the impact of the magnetic field strength as well as the effects of the heat blanketing envelope, initial rotation frequency, NS mass, and core EOS.
Most of our conclusions coincide with the results of previous works on this topic, with one important exception.
We have found a strong dependence of the cooling curves during the rotochemical heating phases on the core EOS, whereas in the literature it was regarded as very weak.
The most likely explanation for this discrepancy is that the authors of the previous works have not considered EOSs with wide enough range of variation of the symmetry energy as a function of density.
\end{abstract}

\section{Introduction} 
\label{sec:Intro}

The composition of baryonic matter inside neutron stars (NSs) is determined by an array of reactions mediated by strong and weak interactions. 
In an unperturbed cold (i.e., with temperatures $T \lesssim 1$~MeV) NS, forward and reverse reactions, e.g., the beta decay of neutrons and electron captures on protons, proceed at the same rates, which means that the NS is chemically equilibrated and its composition does not evolve over time, while escaping neutrinos cool the star.

If a star is slowly compressed, e.g., by accretion or due to spin-down, the local density of matter inside it increases.
Reactions mediated by the strong interaction are fast enough to remain in equilibrium.
In contrast, weak reactions are relatively slow and may not be able to adjust the composition to the state corresponding to weak equilibrium at the increased density.
As a result, the matter moves out of weak\,\footnote{In the following, we will use the terms ``weak'' and ``beta'' as synonyms.} equilibrium. 
This gives rise to non-equilibrium weak reactions whose forward and reverse rates are no longer equal.   
Such non-equilibrium reactions generate entropy and, thus, can heat up the star.
The first calculations of non-equilibrium reaction rates were performed by \cite{Haensel_1992} and employed in simulations of the accretion-induced collapse of NSs into black holes \citep{GE_1993}.

\looseness=-1
The same principle, applied to the spin-down of rotating NSs, is called \emph{rotochemical heating}.
It was introduced in a pioneering work by \cite{Reisenegger_1995} using a schematic model and has since been studied intensively for thirty years. 
\cite{Reisenegger_1997} added the effects of nucleon superfluidity in a simplified way;
\cite{Fernandez_2005} developed a rigorous treatment of the non-superfluid case; \cite{Reisenegger_2006} refined it to include the effects of electric fields that maintain electric neutrality;
\cite{Petrovich_2010} extended the model to properly take into account superfluidity in the approximation of constant and uniform gaps, while \cite{Gonzalez_2015} relaxed this assumption and considered anisotropic and density-dependent gaps.
\cite{Kantor_2021} and \cite{Rodriguez_2026b} examined the impact of magnetic fields in the cores of NSs on rotochemical heating.
\cite{Kantor_2021} also explicitly considered how preceding accretion can ``charge'' the star with chemical energy and keep it warm on a timescale of 1~Gyr.
\cite{Rodriguez_2026a} compared rotochemical heating with other reheating mechanisms to find the best description of some of the old but relatively warm ordinary and millisecond pulsars.
The case of rotochemical heating in quark stars was investigated by \cite{Cheng_1996}.
Research on crust compression was also conducted.
\cite{Iida_1997}\ investigated the rotochemical heating in catalyzed crusts, while \cite{Gusakov_2015} focused on accreted crusts.

\looseness=-1
The aforementioned articles rely on the following set of simplifying assumptions:
1) the mechanical structure is spherically symmetric (1D);
2) the redshifted internal temperature is constant throughout the whole volume of star;
3) also, the redshifted deviations from beta equilibrium are constant throughout the whole volume of star;
4) changes in composition due to spin-down can be computed in the slow rotation approximation \citep{Hartle_1967,Hartle_1968}. 
In this manner, the problem is reduced to a system of two (or three, if muons are taken into account) coupled differential equations that describe the time evolution of the redshifted internal temperature and the redshifted deviation(s) from beta equilibrium.
This allows one to perform analytically a significant part of the calculations, making this approach very efficient.

However, rotating NSs, especially millisecond pulsars, are not spherically symmetric. 
In this paper, we present pioneering simulations of rotochemical heating in full General Relativity in axially symmetric (2D) models without relying on the slow rotation approximation.
We track the thermal, mechanical, and chemical evolution of spinning-down NSs, taking into account the 2D (radial and angular) distributions of temperature, composition, and mechanical structure.
As a first step, we employ the simplest model: only nucleonic composition, only non-equilibrium modified Urca processes  as the main heating/cooling mechanism, and no superfluidity.
NSs with exotic degrees of freedom, superfluidity, and direct Urca processes will be considered elsewhere.

Finally, it is worth noting that rotochemical heating is not the only effect that may occur in spinning down rotating NSs.
Other possible \emph{thermorotational} effects include the redistribution of superfluid pairing gaps and the opening of direct Urca processes \citep{Negreiros_2013,Negreiros_2017}.

The remainder of the paper is organized as follows.
In Sect.~\ref{sec:Model}, we describe the ingredients of our model: dipolar magnetic braking, NSs' mechanical structure and its evolution, non-equilibrium weak reactions associated with heating and chemical evolution, and the core equation of state (EOS). 
Sect.~\ref{sec:Results} is dedicated to our results. 
In Sect.~\ref{sec:Concl}, we draw conclusions and comment on possible further developments.
Throughout this work, we will refer to \citep{Beznogov_ApJ_2023} as Paper~I.

\section{Model} 
\label{sec:Model}

To study rotochemical heating, we employ the simplest model of dipolar magnetic braking with a constant magnetic field and a constant moment of inertia.
In this case, one has (e.g., \citealt{Manchester_1977}):
\begin{equation}
	\frac{d \Omega}{d t} = -f_\theta \frac{B_\mathrm{dip}^2 R^6}{I c^3} \Omega^3 = -K_0 B_\mathrm{dip}^2 \Omega^3
	\label{Eq:Omega_t}
\end{equation}
where $f_\theta \approx 1$ is a numerical factor depending on the geometry of the magnetic field and the model of the magnetosphere, $B_\mathrm{dip}$ is the dipolar magnetic field strength, $R$ is the radius of the star, $I$ is its moment of inertia, and $c$ is the speed of light.   
Substituting fiducial values, we obtain $K_0 = 3.56 \times 10^{-41}$. 
Here, we are only interested in the quadratic dependence of $d\Omega/dt$ on the magnetic field strength.
So, we will use $K_0 = 3.56 \times 10^{-41}$ for all our calculations regardless of its dependence on the field geometry, star mass, EOS, etc.
Note also that our model can handle arbitrary functions $\Omega(t)$ as long as the change of the frequency is not too fast. 

Since we focus on long-term thermal evolution processes that happen on a timescale of hundreds of thousands of years or more, we can safely treat the evolution of the mechanical structure and metric potentials as a sequence of hydrostatic equilibria computed for the respective value of $\Omega$. 
Similarly, it is safe to neglect any \emph{dynamical} variations of the metric potentials caused by the mechanical restructuring of the star.
Thus, at each timestep we employ the methodology described in Paper~I for the metric, mechanical structure, and thermal evolution equations. 

\looseness=-1
Here we briefly list the key aspects and refer the interested reader to Paper~I for more details.
We employ the \texttt{LORENE} library\,\footnote{\url{https://gitlab.in2p3.fr/lorene/Lorene}} \citep{LORENE_2016} to calculate the mechanical structure and the metric potentials of an axisymmetric NS rigidly rotating with an arbitrary frequency $\Omega$.
No slow rotation approximations are involved here.
We use the same quasi-isotropic coordinate system as the one in \texttt{LORENE} and the same set of assumptions about the motion of the matter inside the star.
Specifically, the motion of the matter is assumed to be
(i) axisymmetric [with respect to the rotational axis],
(ii) stationary, 
(iii) purely azimuthal [i.e., no convection, no diffusion, no meridional currents; we also neglect all mass flows induced by heat transport]. 
Finally,
(iv) the EOS of the matter is barotropic. 
We will comment on the last point in Sect.~\ref{ssec:EOS}.

The line element (interval) in \texttt{LORENE} notation and quasi-isotropic coordinates can be written as:
\begin{align}
	\begin{split}
		ds^2 =  g_{\mathit{\alpha \beta}} d x^\alpha d x^\beta = -&N^2 c^2 d t^2 + A^2 \left(d r^2 + r^2 d \theta^2 \right) + \\
		&B^2 r^2 \sin^2 \theta \left(d \varphi - N^\varphi d t \right)^2  .
	\end{split}
	\label{Eq:interval}
\end{align}
where $g_{\mathit{\alpha \beta}}$ is the metric tensor ($\alpha, \beta =$ 0, 1, 2, and 3), $x^\alpha = \left(ct,r,\theta,\varphi \right)^\mathrm{T}$ is the 4-coordinate, $N$, $A$, $B$ and $N^\varphi$ are four metric functions that depend on $r$ and $\theta$ but do not depend on $t$ and $\varphi$ due to stationarity (ii) and axisymmetry (i), respectively.
Note that $N$, $A$, $B$ are dimensionless while $N^\varphi$ has the dimension of frequency.

In the following subsections, we will describe various key aspects of our model.
We only consider NSs made of nucleonic $\mathit{npe\mu}$ matter.

\subsection{Mechanical structure and its evolution} 
\label{ssec:Structure}

In the absence of accretion and/or mass loss, the number of baryons inside a star is conserved.
This is exactly the situation considered in this work.
However, unlike the case of spherically symmetric stars, the enclosed baryon number cannot be used directly as a (radial) Lagrangian variable. 
A more elaborate approach is needed.
One has to specify the iso-contours that enclose a fixed number of baryons and track how these contours evolve as the star spins down.
\cite{Fernandez_2005} demonstrated that in a rigidly rotating perfect-fluid star the iso-surfaces of pressure, energy density and other thermodynamic quantities coincide even for non-barotropic EOSs and, consequently, enclose a fixed number of baryons throughout the evolution of the star.
In other words, as the star spins down, the isobars deform and the pressure associated with them increase, yet they still enclose the same number of baryons.
This serves as a basis for an invariant (Lagrangian) grid.
 
Thus, for a given EOS and baryonic mass of an NS, we begin at $t = 0$ with some initial rotation frequency $\Omega_0$.
We use a uniform angular grid $\{\theta_j, j = 0,1, \ldots, j_\mathrm{max} \}$ that coincides with the \texttt{LORENE} angular grid and does not change over time. 
The radial grid $\{r_i, i = 0,1, \ldots, i_\mathrm{max} \}$ is initialized at $t = 0$ similarly to the ``standard'' \texttt{NSCool} \citep{NSCool} radial grid.
After the  $\{r_i,\theta_j\}$ grid is set up, we first compute the isobars $r_i(\theta)$ for each radial grid point $r_i$ based on the value of pressure at $\theta = 0$.
Then we compute the corresponding enclosed baryon number in each shell (the mechanical structure is symmetric w.r.t. the equatorial plane):
\begin{align}
	N_i = 4\pi\int_0^{\frac{\pi}{2}} \sin \theta d\theta \int_{r^i_\mathrm{min}(\theta)}^{r^i_\mathrm{max}(\theta)} n_\mathrm{B}\,\Gamma A^2 B\, r^2 dr,
	\label{Eq:EnclosedNBar}
\end{align}
where $ i = 1, \ldots, i_\mathrm{max}$, $n_\mathrm{B}$ is the baryon number density, $r^i_\mathrm{min}(\theta)$ and $r^i_\mathrm{max}(\theta)$ define the isobars between which we compute the number of baryons, 
\begin{align}
	\Gamma = \left[1-\left(\frac{U}{c}\right)^2\right]^{-\frac{1}{2}}
	\label{Eq:GammaLorentz}
\end{align}
is the Lorentz-factor, and
\begin{align}
	U = B\,r \sin \theta (\Omega - N^\varphi)/N
	\label{Eq:Vel}
\end{align}
is the fluid element velocity. 
Obviously,  $r^{i=1}_\mathrm{min}(\theta) \equiv 0$ and  $r^i_\mathrm{max}(\theta)  \equiv r^{i+1}_\mathrm{min}(\theta)$.

Equation~\eqref{Eq:EnclosedNBar} gives us our Lagrangian grid $\{N_i, \theta_j\}$ and a mapping to the $\{r_i,\theta_j\}$ grid, which is the one actually used for all thermal and chemical evolution calculations.
At each timestep, Eq.~\eqref{Eq:EnclosedNBar} is solved numerically to obtain the new coordinates $r^i_\mathrm{min}(\theta)$ and $r^i_\mathrm{max}(\theta)$ of the isobars enclosing $N_i$ baryons.
Details of the numerical implementation are provided in Appendix~\ref{App:Numeric}. 
See also Sect.~\ref{ssec:EOS} and~\ref{ssec:ThermEvol}.

\subsection{Non-equilibrium weak reactions} 
\label{ssec:Reactions}

As discussed in the Introduction, the compression of the star due to spin-down causes the matter to deviate from beta equilibrium.
This modifies the neutrino emission rates.
By far, the dominant mechanism of neutrino emission from a non-superfluid \emph{canonical} mass ($M_G = 1.4~\Msun$) NS is the modified Urca process (see, e.g., \citealt{Pethick_1992,Yakovlev_2001,Page_2006} and also Sects.~\ref{ssec:EOS} and~\ref{ssec:ThermEvol} below).
Thus, we only consider the impact of deviations from the beta equilibrium on the modified Urca emissivity.
All other neutrino emission mechanisms in the core and in the crust are assumed to operate as in beta-equilibrium matter.

The modified Urca reactions can be written as:
\begin{align}
	&N + n \to N + p + l + \tilde{\nu}_l, 
	\label{Eq:beta_decay} \\
	&N + p + l  \to N + n + \nu_l,
	\label{Eq:beta_capture}
\end{align}
where $N$ is a spectator nucleon (neutron $n$ or proton $p$), $l$ denotes a lepton (electron $e$ or muon $\mu$), and $\nu_l$ is a corresponding (anti-)neutrino.
In beta equilibrium, the beta decay \eqref{Eq:beta_decay} and the beta capture \eqref{Eq:beta_capture} reaction rates are identical (principle of detailed balance), thus yielding the conditions of beta equilibrium (chemical potentials include rest-masses):
\begin{align}
	\mu_n = \mu_p + \mu_l. 
	\label{Eq:beta_eq} 
\end{align}
We assume that neutrinos can leave the system unhindered, so their chemical potentials are zero.
For the temperatures we are interested in, i.e., $k_\mathrm{B} T \ll 1$~MeV, this is fully justified. 
Note that reactions~\eqref{Eq:beta_decay} and~\eqref{Eq:beta_capture} are, in fact, not the exact inverses of each other. 
This additionally limits the domain of validity of condition~\eqref{Eq:beta_eq}, see \citep{Alford_2018,Alford_2024} for more details.
Nevertheless, at $k_\mathrm{B} T \ll 1$~MeV it is valid, which is sufficient for our purposes.

One can see that by applying beta equilibrium conditions for both electrons and muons, one automatically gets $\mu_e = \mu_\mu$ without the need to explicitly take into account leptonic reactions. 

If the system is out of beta equilibrium, the reactions~\eqref{Eq:beta_decay} and~\eqref{Eq:beta_capture} have unequal rates.
We can quantify the deviations from equilibrium by introducing chemical potential imbalances
\begin{align}
	\Delta \mu_l = \mu_n - \mu_p - \mu_l. 
	\label{Eq:delta_mu} 
\end{align}
We assume that $\Delta \mu_l \ll \mu_n, \mu_p, \mu_l$, but the imbalances can be larger than $k_\mathrm{B}T$. 
Following the derivation of  \cite{Haensel_1992,Reisenegger_1995,Yakovlev_2001}, for non-superfluid matter, one can write:
\begin{align}
	Q_\mathrm{noneq} = Q_\mathrm{eq} F(\xi), \quad \Delta \Gamma = b Q_\mathrm{eq} H(\xi) \frac{\xi}{k_\mathrm{B}T},
	\label{Eq:MURCA_noneq} 
\end{align}
where $Q_\mathrm{noneq}$ is the emissivity of the modified Urca (any branch) process out of beta equilibrium, $Q_\mathrm{eq}$ is the corresponding emissivity in beta equilibrium, $\xi = \Delta \mu/(k_\mathrm{B} T)$ is the dimensionless chemical potential imbalance, $b$ is a numerical constant, $\Delta \Gamma$ is the net rate of conversion of neutrons into protons (in case $\xi > 0$); the total baryon number density $n_\mathrm{B} = n_n + n_p$ is obviously conserved. 
Finally, the functions $F(\xi)$ and $H(\xi)$ determine the impact of deviations from the chemical equilibrium.
It should be noted that in non-superfluid matter, these functions and the general decomposition~\eqref{Eq:MURCA_noneq} are independent of any employed model of nuclear matter (NM) since they are based solely on the phase-space decomposition \citep{Yakovlev_2001}. 
On the other hand, the equilibrium emissivity $Q_\mathrm{eq}$ depends strongly on the model of NM and theoretical approaches used to compute it, see, e.g., \citep{Shternin_2018,Suleiman_2023,Alford_2024_nwa}.

The functions $F(\xi)$ and $H(\xi)$ were calculated analytically by \cite{Reisenegger_1995}.
In the notation of \citep{Yakovlev_2001}, we have
\begin{align}
	&b = \frac{14\,680}{11\,513 \pi^2}, 	\label{Eq:F&H}  \\
	&F(\xi) = 1 + \frac{22\,020 \xi^2}{11\,513 \pi^2} + \frac{5670 \xi^4}{11\,513 \pi^4} + \frac{420 \xi^6}{11\,513 \pi^6} + \frac{9\xi^8}{11\,513 \pi^8}, \nonumber \\
	&H(\xi) = 1 + \frac{189 \xi^2}{367 \pi^2} + \frac{21 \xi^4}{367 \pi^4} + \frac{3 \xi^6}{1835 \pi^6}. \nonumber
\end{align}
In beta equilibrium one has $\xi = 0$, $F(0) = H(0) = 1$ and $Q_\mathrm{noneq} = Q_\mathrm{eq}$, while $\Delta \Gamma = 0$.

Once we have the net reaction rates $\Delta \Gamma$, we can formulate the equations of chemical evolution of the matter inside the NS core. 
Those equations are:
\begin{align}
	\frac{d y_e}{d t}   = \frac{\Delta \Gamma_e}{n_\mathrm{B}}, \quad
	\frac{d y_\mu}{d t} = \frac{\Delta \Gamma_\mu}{n_\mathrm{B}},
	\label{Eq:chem_evol} 
\end{align}
where $y_l = n_l/n_\mathrm{B}$ is the (number) fraction of electrons and muons.
These must be be solved at each timestep along with the charge neutrality condition $y_p = y_e + y_\mu$.

\looseness=-1
Non-equilibrium weak reactions not only modify the composition, but also impact the thermal evolution. 
The first aspect here is straightforward: the neutrino emissivities are modified from $Q_\mathrm{eq}$ to $Q_\mathrm{noneq}$.
The second aspect is due to the fact that the relaxation to beta equilibrium~\eqref{Eq:chem_evol} is an irreversible thermodynamic process that generates entropy and thus heat \citep{Haensel_1992,GE_1993,Reisenegger_1995}. 
For $\Delta \mu_l \ll \mu_n, \mu_p, \mu_l$, the heating associated with this process is simply 
\begin{align}
	Q_{\mathrm{heat},l} = \Delta \Gamma_l \Delta \mu_l.
	\label{Eq:chem_heat} 
\end{align}
Then the total rate of change of the thermal energy per unit volume due to the non-equilibrium modified Urca process can be written as:
 \begin{align}
 	Q_\mathrm{tot} = \Delta \Gamma \Delta \mu - Q_\mathrm{noneq} = Q_\mathrm{eq} 
 	\left[b \xi^2 H(\xi) - F(\xi) \right],
 	\label{Eq:heat_tot} 
 \end{align}
where we sum over the electronic and muonic branches of modified Urca.
If $\xi$ exceeds 5.5, the heating overcomes neutrino cooling. 
In our context, this is the rotochemical heating.

Finally, we have employed equilibrium modified Urca rates from \cite{Yakovlev_2001} for $Q_\mathrm{eq}$. 
Although those are old, they are simple and fast numerically.
Modern approaches like \citep{Shternin_2018,Suleiman_2023,Alford_2024_nwa} are significantly more computationally involved and might be difficult to incorporate into our thermal evolution code.

\subsection{Equations of state} 
\label{ssec:EOS}

The concept of beta equilibrium is extremely important for NSs.
Chemical equilibrium determines the composition of a star. 
The condition~\eqref{Eq:beta_eq} applied to a particular EOS of dense matter gives the dependence of the proton fraction $y_p = n_p/n_\mathrm{B}$ on $n_\mathrm{B}$. 
This dependence, in turn, determines whether direct Urca processes of neutrino emission \citep{Boguta_1981,Lattimer_1991} are allowed inside the NS cores.
The presence or absence of direct Urca processes is crucial for the thermal evolution of isolated NSs and transiently accreting NSs in low-mass X-ray binaries (e.g., \citealt{Yakovlev_2004,HPY07}).

\begin{figure}
	\includegraphics[]{"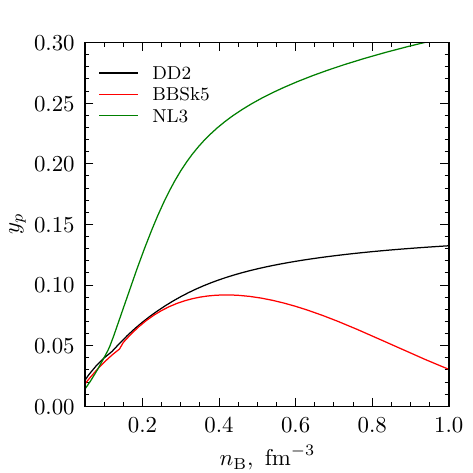"}
	\caption{The proton fraction as a function of baryon number density in NS matter in beta equilibrium for all three considered core EOSs.}
	\label{Fig:Yp_NS}
\end{figure}

The modern consensus is that it should operate in sufficiently massive NSs in order to explain the observational data on the coldest known isolated NSs and, especially, the accreting ones (see., e.g., \citealt{Beznogov_2015a,Beznogov_2015b,Brown_2018,Potekhin_2019}).
On the other hand, direct Urca is not expected to operate in canonical-mass NSs as otherwise, many of the observed ``warm'' stars could not be explained.
Thus, it is usually considered that the threshold NS mass where direct Urca begins to operate is $M_\mathrm{D} \gtrsim 1.5~\Msun$ \citep{Klahn_2005,Beznogov_2015b} and therefore exceeds the canonical NS mass of $1.4~\Msun$. 
Alternatively, one can  argue that the threshold mass is lower, but for canonical mass NSs the direct Urca is completely suppressed by nucleon superfluidity, while for more massive and denser  NSs this suppression becomes weaker and finally disappears.
Such a scenario is possible because superfluid gaps as functions of density are bell-shaped; see, e.g., \citep{Ding_2016} for modern gap calculations.

Regardless, we focus here exclusively on modified Urca processes and disregard direct Urca processes whether they are kinematically allowed or not.
     
From the point of view of an EOS of dense matter, $y_p$ in beta equilibrium is determined by the density dependence of the symmetry energy and from Sect.~\ref{ssec:Reactions} it follows that rotochemical heating is centered on changes of $y_p$ and the corresponding energy release.
Thus, we have chosen three core EOSs that provide vastly different density dependencies of the symmetry energy: DD2~\citep{DD2}, BBSk5~\citep{RB_2025,RB_2026}, and NL3~\citep{NL3}.
It should be noted that the slope of the symmetry energy for the NL3 EOS is 118.5~MeV, which is no longer consistent with current constraints.
However, since one of our goals was to assess the impact of the core EOS on rotochemical heating, we decided to use the NL3 EOS as an extreme case scenario.

Figure~\ref{Fig:Yp_NS} demonstrates the dependence of the proton fraction in NS matter in beta equilibrium on the baryon number density for all three considered core EOSs. 
One can clearly see that the EOSs indeed have vastly different $y_p(n_\mathrm{B})$ dependencies.
The NL3 EOS allows for direct Urca at just $\approx$1.5 nuclear saturation density, but, as mentioned above, we disregard it.

As we investigate long-term thermal evolution and typical temperatures are much less than 1~MeV, we can safely use zero-temperature core and crust EOSs, see a detailed discussion in Paper~I. 
However, unlike the situation in Paper~I, here we have deviations from beta equilibrium, and our EOSs have to accommodate these. 
So, we employed the following strategy.
Since we did not consider the impact of rotochemical heating on the crust, we used only one crustal EOS (see below), and it was taken at beta equilibrium for all computations.
The core EOSs were taken at beta equilibrium for calculations of mechanical structure since
(a) \texttt{LORENE} requires barotropic EOS,
(b) the pressure vs energy-density dependence ($P(\rho)$) is virtually unaffected by the small deviations from chemical equilibrium.
Even for $\Delta \mu \approx 10$~MeV, the deviations of pressure do not exceed a few percent, but as we show in Sect.~\ref{ssec:Results_Chemical}, in our simulations $\Delta \mu$ is much less than 1~MeV. 

For the core composition, we employed 3D versions of the core EOSs, i.e., with composition depending on $(n_\mathrm{B},y_p,y_\mu)$. 
This allowed us to accommodate departures from beta equilibrium for both electrons and muons. 
Naturally, in the case $\Delta \mu_e = \Delta \mu_\mu = 0$ those 3D EOSs reduce back to beta equilibrium barotropic EOSs that were used for mechanical structure calculations.

For the crust EOS we had to employ a ``hybrid'' approach due to the specific requirements of \texttt{LORENE}.
Being a spectral code, \texttt{LORENE} on one hand works with very high numerical accuracy, but, on the other hand, it is very sensitive to the smoothness of the EOS.
Any irregularities, which frequently happen in the crust-core transition region, propagate into the mechanical structure.
For a fixed rotation frequency, as in Paper~I, this is not a big issue.
But the grid mapping procedure described in Sect.~\ref{ssec:Structure} requires numerical integration and equation-solving and, consequently, greatly amplifies any irregularities.
As a result, in Paper~I we could afford to work with \texttt{LORENE} data computed with errors on GRV identities as high as $\sim 5\times 10^{-5}$. 
Here, we had to keep these errors below $10^{-7}$, otherwise we could not follow the evolution of the mechanical structure during the spin-down.
Thus, the composition of the crust was taken directly from \citep{Haensel_1989} and \citep{Negele_1973}, similarly to Paper~I.
The $P(\rho)$ dependence in the crust had to be smoothed, especially in the crust-core transition region.
To do that, we employed \texttt{make\_splrep} from \texttt{scipy.interpolate}.
The smoothing factor was between 0.02 and 0.05 depending on the core EOS.
We used smoothing only in a narrow density range around the crust-core transition density.

\subsection{NS models \& long-term thermal evolution \& other remarks} 
\label{ssec:ThermEvol}
  
We consider the long-term thermal evolution of spinning-down non-accreting NSs of baryonic masses $M_\mathrm{B} = 1.6~\Msun$ and 2.3~$\Msun$.
This is roughly equivalent to gravitational masses $M_\mathrm{G} = 1.4~\Msun$ and 2.0~$\Msun$, respectively.
The first mass corresponds to a canonical NS.
Note that the gravitational mass decreases as the star spins down, while the baryonic mass is conserved.
For a detailed description of the concept of long-term cooling, see Paper~I and the references therein.

\looseness=-1
The effects of superfluidity are disregarded in this work.
We also neglect the impact of magnetic fields on thermal evolution because the magnetic fields of interest for rotochemical heating are of the order of $10^{8}-10^{9}$~G \citep{Reisenegger_1995}, which is much lower than the typical values for NSs.
Magnetic fields of this strength have almost no effect on thermal evolution \citep{Yakovlev_2004,Potekhin_2015}.       

The standard approach for long-term thermal evolution of NSs is to split a star into the star itself and a heat-blanketing envelope, which is treated separately and in a time-independent manner by means of the so-called ``$T_\mathrm{s} - T_\mathrm{b}$'' relations (see, e.g., Paper~I, \citealt{Yakovlev_2004,HPY07,Beznogov_2021}).
Usually, the splitting is performed at the boundary density $\rho_\mathrm{b} = 10^{10}$~g/cm$^3$.
In our case, however, despite smoothing the crustal EOS as described in Sect.~\ref{ssec:EOS}, the calculations of the time evolution of the mechanical structure of the star's outer layers were not numerically stable enough.
Thus, we had to place the boundary density deeper, at the neutron drip point $\rho_\mathrm{b} = 4 \times 10^{11}$~g/cm$^3$.
Correspondingly, we have chosen an envelope model by \cite{Potekhin_2003} because it has the same value of $\rho_\mathrm{b}$.
Since the rotochemical heating scenario is most applicable to millisecond pulsars that might have been spun up by accretion, we performed most of the calculations with a fully accreted envelope ($\Delta M = 10^{-8}~\Msun$).
For comparison, we also performed some calculations with a non-accreted (iron) envelope ($\Delta M = 10^{-31}~\Msun$).
   
We start our simulations with an initial rotation frequency of 700 or 800~Hz and an initial redshifted temperature of $2 \times 10^9$~K.
We assume that initially the weak reactions are fast enough so that the star is in beta equilibrium.
Then we allow the system to spin down following Eq.~\eqref{Eq:Omega_t}. 

In beta equilibrium $\mu_\mu = \mu_e$ (see Sec~\ref{ssec:Reactions}) and muons appear as soon as the electron chemical potential exceeds the rest-mass of the muon. 
In the absence of chemical equilibrium, $\mu_\mu$ is no longer equal to $\mu_e$, which complicates the tracking of the muon onset threshold.
Although lepton decay processes $e + l \to \mu + l + \nu_e + \bar{\nu}_\mu$ and $\mu + l \to e + l + \nu_\mu + \bar{\nu}_e$ exist \citep{Alford_2010} and can, in principle, equalize $\mu_e$ and $\mu_\mu$, they are orders of magnitude slower than the modified Urca processes.
Moreover, their reaction rates are known only in the limit $\mu_e - \mu_\mu \ll k_\mathrm{B} T$, and we need the opposite limit, see the discussion in \citep{Kantor_2021}.
Thus, we disregard these reactions and consider that muons cannot appear in Lagrangian zones where there were no muons at $t = 0$.
This affects only 1-2 radial grid points and has absolutely no impact on the results.

\section{Results} 
\label{sec:Results}

%
\begin{figure*}
	\centering
	\includegraphics[]{"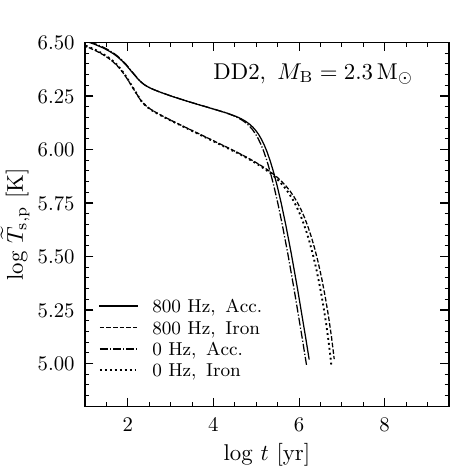"}
	\includegraphics[]{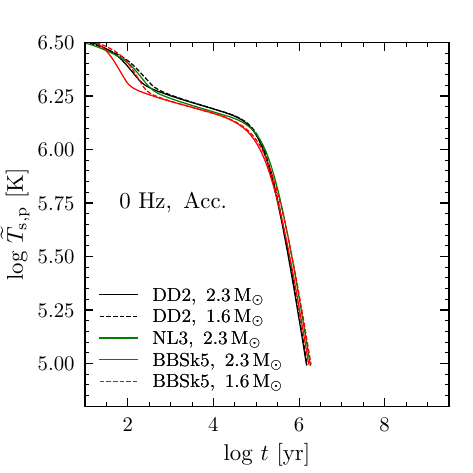""}
	\caption{
		The cooling curves (redshifted surface temperature at the pole vs age) for NSs with constant rotation frequency and for non-rotating NSs (i.e., no rotochemical heating). 
		Different heat blanketing envelopes and rotation frequencies are plotted in the left panel, while various EOSs of the core and baryonic masses are plotted in the right panel. 
		``Acc.'' stands for fully-accreted envelope. }
	\label{Fig:Cool_NoRT}
\end{figure*}
\begin{figure*}
	\centering
	\includegraphics[]{"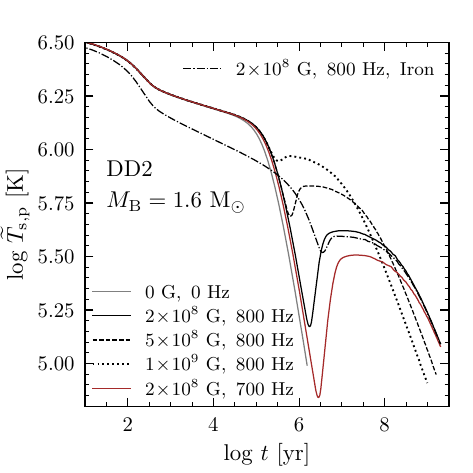"}
	\includegraphics[]{"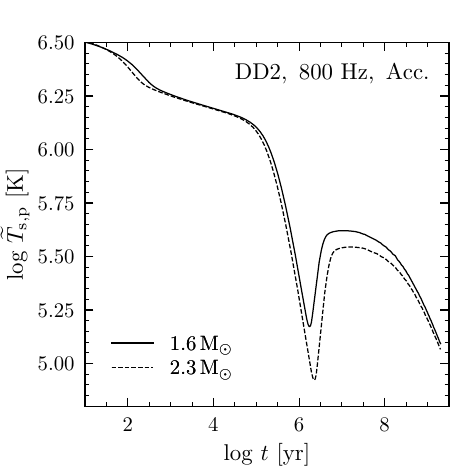"}
	\caption{
		The cooling curves with rotochemical heating. 
		The impacts of the magnetic field strength, the initial rotation frequency, and the envelope composition are shown in the left panel, while the impact of the NS mass is demonstrated in the right panel. 
		DD2 EOS in all cases. All curves correspond to fully-accreted envelopes except for the dash-dotted curve that corresponds to an iron envelope.}
	\label{Fig:Cool_RT}
\end{figure*}
\begin{figure}
	\centering
	\includegraphics[]{"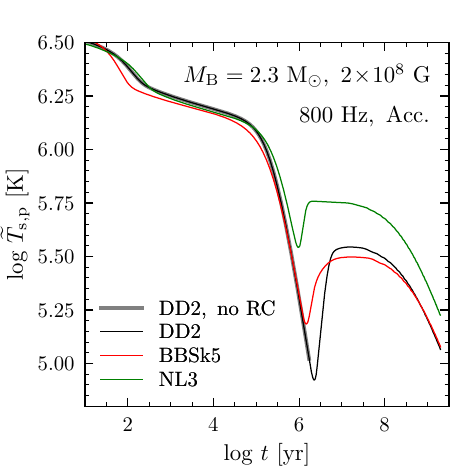"}
	\caption{
    The cooling curves with rotochemical heating demonstrating the effect of the core EOS for NSs with identical mass, magnetic field, initial rotation frequency, and envelope composition. The thick gray curve corresponds to an NS with a fixed rotation frequency, which does not experience rotochemical heating (RC).}
	\label{Fig:Cool_EOS}
\end{figure}
\begin{figure*}
	\centering
	\includegraphics[]{"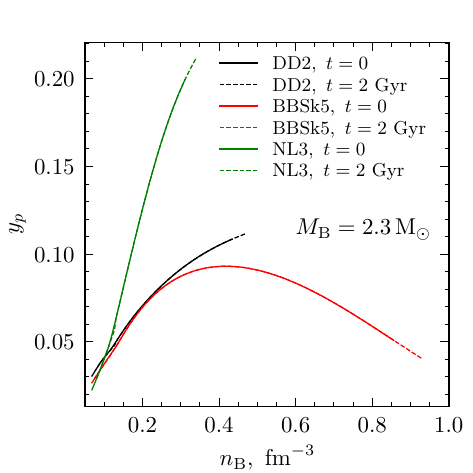"}
	\includegraphics[]{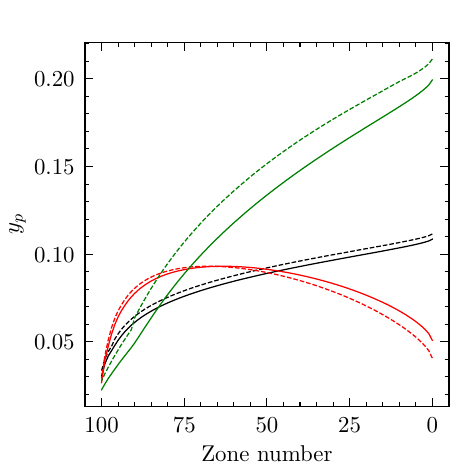}
	\caption{
		The proton fraction in the cores of NSs with different core EOSs, shown at the beginning of the simulations and after 2~Gyr.
		Parameters correspond to Fig.~\ref{Fig:Cool_EOS}.
		The left panel demonstrates $y_p$ as a function of $n_\mathrm{B}$, while the right panel demonstrates it as a function of the invariant (Lagrangian) zone number.
		In both panels, the crust-core boundary is on the left and the center of the star is on the right.}
	\label{Fig:Yp}
\end{figure*}
\begin{figure*}
	\centering
	\includegraphics[]{"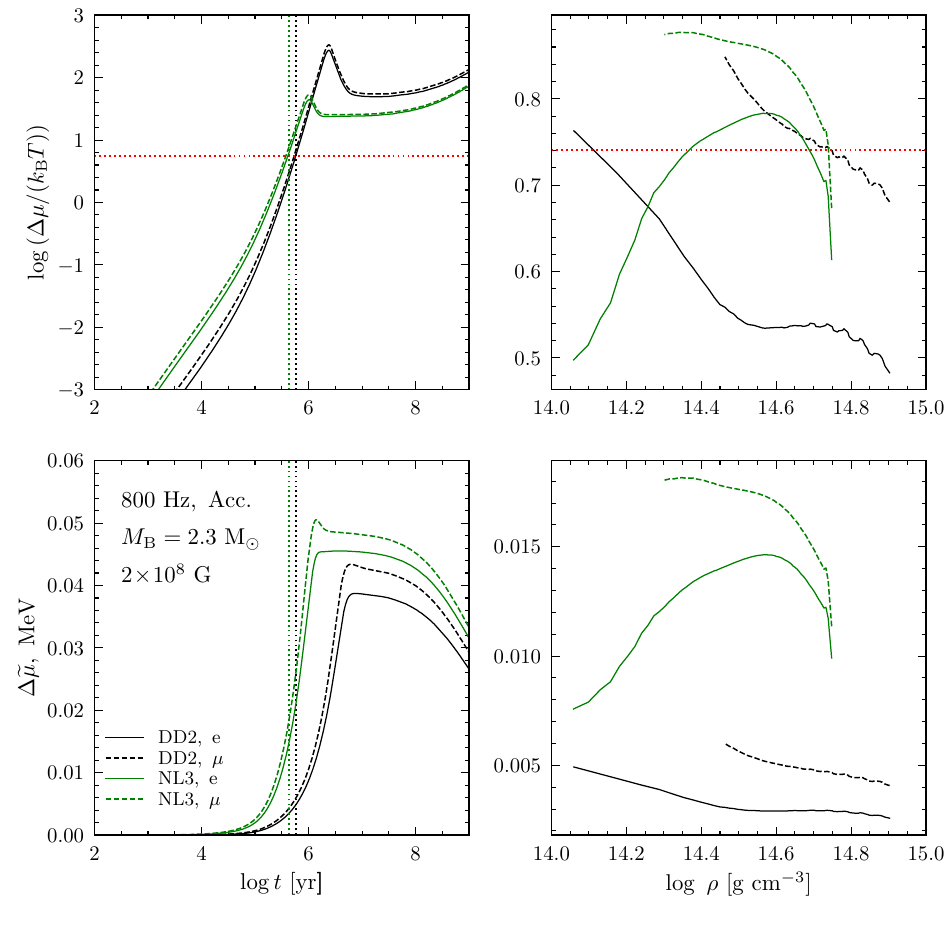"}
	\caption{
		The deviations from chemical equilibrium for electrons and muons, shown in terms of $\xi_l = \Delta \mu_l/(k_\mathrm{B}T)$ (upper panels, $l=e,\mu$) and $\Delta \widetilde{\mu}_l$ (redshifted, lower panels) as a function of NS age (left panels) and density (right panels).
		The maximum deviation across the core is demonstrated as a function of time.
		For the times marked by vertical dotted lines (black and green), the deviations are shown as a function of density.}
	\label{Fig:delta_mu}
\end{figure*}

We start with a quick  review of the long-term thermal evolution of non-rotating NSs and NSs rotating with constant frequency.
We  compare NSs with various masses, core EOSs, and heat-blanketing envelopes.
These results are well-known.
We provide them here to serve as a basis for comparison with the results that take rotochemical heating into account.
In Paper~I we demonstrated that the temperature at the star's pole is least affected by rotation.
Since we investigate stars with changing frequency, it is logical to present the cooling curves by plotting the redshifted temperature at the pole as a function of the star's age.

The left panel in Fig.~\ref{Fig:Cool_NoRT} demonstrates the cooling curve for NSs with the DD2 EOS of the core and baryonic mass $M_\mathrm{B} = 2.3~\Msun$. 
By comparing 800~Hz curves with the corresponding (i.e., with the same envelope) 0~Hz curves, one can see that, indeed, the temperature at the pole is almost unaffected by rotation.
Conversely, by comparing the curves with the same frequency and different envelopes, the impact of the envelope is clearly seen.
In the right panel, one can see the impact of the core EOS and the star mass on the cooling curves. 
Both factors have a limited influence on the cooling curves. 
Most of the differences occur during the crust-core thermalization epoch (i.e., around the age $t \sim 100$~yr) and can be explained by the fact that different EOSs and different masses result in different thicknesses of the crust.
The thinner the crust, the shorter the heat diffusion timescale and the faster it thermalizes with the core.
Since the core is colder than the crust as a result of more intensive neutrino emission, the temperature in the crust (and on the surface) decreases faster for thinner crusts. 
We can clearly see this by comparing the two DD2 curves (black) with different masses.
A higher mass results in a thinner crust, and the temperature at $t \sim 100$~yr drops faster. 
The same is observed for the two BBSk5 curves (red). 
Then, the BBSk5 EOS is softer than the DD2 EOS.
Thus, the BBSk5 NS has a thinner crust than the DD2 NS of identical baryonic mass.
Consequently, the temperature of BBSk5 stars at $t \sim 100$~yr drops faster than that of the corresponding DD2 stars. 
Note also that the conclusions shown above will change significantly if superfluidity is taken into account.
  
\subsection{Thermal Evolution} 
\label{ssec:Results_Thermal}
   
Now, let us move on to our main results for rotochemical heating.
The left panel in Fig.~\ref{Fig:Cool_RT} shows the cooling curves for an NS with the DD2 EOS and $M_\mathrm{B} = 1.6~\Msun$. 
Various magnetic fields, initial rotation frequencies, and envelope compositions are considered as indicated in the legend.
Let us first discuss the main common features and establish a nomenclature.
For the magnetic fields of interest ($10^8 - 10^9$~G), the effects of rotochemical heating manifest themselves during the photon cooling era, i.e., for $t \gtrsim 300$~kyr; preceding epochs remain unaffected.
One can see three distinct phases of rotochemical heating, which we will refer to as the (initial) heating phase, the quasi-equilibrium phase (plateau), and the (final) cooling phase. 
The quasi-equilibrium phase was introduced and discussed in detail by \cite{Reisenegger_1995} and subsequent works. 
The other two phases do not have generally accepted names, so we proposed the names.

The impact of the magnetic field strength is clearly visible (black solid vs dashed vs dotted curves). 
The stronger the field, the earlier the rotochemical heating effects become visible. 
The quasi-equilibrium state temperature is higher, but its duration is shorter than in cases with lower magnetic fields.
The resulting temperature at $t=2$~Gyr is also lower.
Lowering the initial rotation frequency delays the onset of rotochemical heating, noticeably increases the temperature rise during the initial heating phase, and slightly lowers the temperature during the quasi-equilibrium phase (solid black vs brown curves).
However, ultimately, the curves with different initial rotation frequencies coincide with each other during the final cooling phase.

Moving on, one can assess the effects of the envelope composition by comparing the solid and dash-dotted black curves.
Before the heating started, these two curves demonstrate a crossing at the age of $t \approx 300$~kyr, when a star switches from the neutrino cooling phase to the photon cooling phase. This is a standard behavior for cooling NSs, see the left panel in Fig~\ref{Fig:Cool_NoRT}.
The situation changes as the rotochemical heating starts. 
The star with a non-accreted envelope (dash-dotted black curve) is much warmer at $t \gtrsim 1$~Myr.
It starts to heat up later, and the initial heating phase is much less pronounced than for its fully accreted counterpart (solid black curve). 
Despite this noticeable difference, the temperature of both stars during the plateau phase is almost identical and the curves coincide completely during the final cooling phase.

\looseness=-1
The effect of the NS mass can be inferred from the right panel in Fig.~\ref{Fig:Cool_RT}.
It shows the cooling curves for NSs identical in every aspect except for their masses.
Before the rotochemical heating starts, the curves nearly coincide (cf. the right panel in Fig.~\ref{Fig:Cool_NoRT}).
Some differences in temperature appear during the initial heating phase.
The heavier NS (black dashed curve) starts to heat up later and at a lower temperature.
Then, during the plateau phase, the differences decrease, and, finally, at $t \gtrsim 1$~Gyr the temperature of both NSs becomes almost the same. 

Thus, we can conclude that for a given magnetic field and core EOS (see below), neither the initial rotation frequency nor the envelope composition affects the temperature of the star at the final cooling phase. 
The mass of the star has almost no impact as well. 
This is in line with the conclusions of \cite{Reisenegger_1995,Fernandez_2005}.
Keep in mind, however, that the preceding evolution, especially the initial heating phase and, in some cases, the plateau phase, depends noticeably on these three parameters, as clearly seen in Fig.~\ref{Fig:Cool_RT}.

The dependence of the results on the core EOS is depicted in Fig.~\ref{Fig:Cool_EOS}.
The cooling curves correspond to NSs that only differ in their core EOSs.
The BBSk5 EOS (red curve) results in a somewhat earlier onset of rotochemical heating and a slightly lower temperature during the plateau phase compared to our reference DD2 model (black curve).
For $t \gtrsim 300$~Myr, the temperature for both EOSs is nearly identical.
The results for the NL3 EOS are considerably different. 
The NS with this EOS (green curve) starts heating up even earlier than the one with the BBSk5 EOS.
The NL3 NS stays significantly hotter than both DD2 and BBSk5 NSs during the plateau phase and remains hotter during the final cooling phase, up to the end time of our simulations (2~Gyr).

Unlike our results related to Fig~\ref{Fig:Cool_RT}, which were in line with the results in the literature, this strong EOS dependence is new.
The apparent contradiction with the conclusions of \citep{Fernandez_2005} most likely stems from the fact that \cite{Fernandez_2005} have not considered EOSs with a sufficiently wide range of possible dependencies of the symmetry energy on the density.
Indeed, as discussed in Sect.~\ref{ssec:EOS}, we have deliberately chosen three core EOSs with vastly different dependencies of the symmetry energy on density.
By comparing Figs.~\ref{Fig:Yp_NS} and \ref{Fig:Cool_EOS}, this parallelism is evident. 
The BBSk5 and DD2 EOSs provide comparable proton fractions ($y_p$) and slopes ($\partial y_p/\partial n_\mathrm{B}$) up to $n_\mathrm{B} \approx 0.30 - 0.35$~fm$^{-3}$, while the NL3 EOS already deviates significantly at nuclear saturation density.
The cooling curves match this behavior: the BBSk5 and DD2 curves are not strongly different from each other, while the NL3 curve deviates noticeably from them.
Apart from that, one can see that the bell-shaped dependence of $y_p$ on $n_\mathrm{B}$, corresponding to the BBSk5 EOS, does not, of itself, lead to any peculiar results in terms of the cooling curves.

\subsection{Chemical Evolution} 
\label{ssec:Results_Chemical}

\looseness=-1
Now, let us consider the evolution of the chemical composition.
Fig.~\ref{Fig:Yp} demonstrates the proton fractions in the cores of NSs from Fig.~\ref{Fig:Cool_EOS} at the beginning of the calculation (solid lines) and after 2~Gyr (dashed lines).
In the left panel $y_p$ is represented as a function of $n_\mathrm{B}$.
Since we start from beta equilibrium, it is not surprising that the curves at $t=0$ look exactly like the ones in Fig.~\ref{Fig:Yp_NS}.
Similarly, since we assume (and confirm afterwards, see below) that the deviations from chemical equilibrium are small, it is also logical that the curves at $t=2$~Gyr look like the initial ones but extend to higher densities.

To better understand what is actually happening, we plot $y_p$ as a function of the invariant (Lagrangian) zone number in the right panel of Fig.~\ref{Fig:Yp}. 
Zone number 100 corresponds to the crust-core boundary, and zone number 0 corresponds to the center of the star. 
Here, the curves for $t=0$ and $t=2$~Gyr are distinct and clearly demonstrate changes in $y_p$.
The difference between the representations of $y_p$ in the left panel and in the right panel comes from the fact that we essentially plot $y_p$ for different conditions.
By plotting $y_p$ as a function of the zone number, we trace how the proton fraction changes in a (physical) cluster of matter that slowly becomes denser as the star spins down (see Sect.~\ref{ssec:Structure}).
However, when plotting $y_p$ as a function of $n_\mathrm{B}$, we track the proton fraction of matter that happens to have a given density at a given time.
In other words, for a fixed density, at $t=0$ and $t=2$~Gyr  we are looking at different (physical) clusters of matter. 

To sum up, the proton fraction of a (physical) cluster of matter changes with time, and this change follows the beta equilibrium path very closely.
The proton fraction of matter at a fixed density is almost constant and deviates very little from the initial beta equilibrium value.  
Thus, we conclude that for our conditions and for small deviations from beta equilibrium, non-equilibrium weak reactions are efficient in keeping the matter very close to equilibrium.

\looseness=-1
Some other remarks related to Fig.~\ref{Fig:Yp} are in order here. 
From the left panel, it is evident that the considered EOSs are very different in stiffness.
For stars with identical baryonic masses, the corresponding central densities differ significantly, with the NL3 EOS being the stiffest and the BBSk5 EOS the softest.
The peculiarity of the BBSk5 EOS (red curves) is revealed on the right panel: at high zone numbers (i.e., low densities) $y_p$ increases with time, while at low zone numbers (i.e., high densities) $y_p$ decreases with time. 
For the two other EOSs $y_p$ always increases with time.
This outcome is a direct consequence of the bell-shaped $y_p(n_\mathrm{B})$ dependence shown in Fig.~\ref{Fig:Yp_NS}.  

Next, we focus on the deviations from beta equilibrium plotted in Fig.~\ref{Fig:delta_mu} for the DD2 (black) and NL3 (green) EOSs for the conditions corresponding to those in Fig.~\ref{Fig:Cool_EOS}.
The upper panels demonstrate the logarithm of the dimensionless chemical potential imbalance $\xi_l = \Delta \mu_l/(k_\mathrm{B}T)$ with $l=e,\mu$, while the lower panels show the redshifted deviation from the beta equilibrium $\Delta \widetilde{\mu}_l$.
The red dotted horizontal line on the upper panels corresponds to $\xi = 5.5$.
As mentioned in Sect.~\ref{ssec:Reactions}, at this value, the heating due to non-equilibrium reactions balances the neutrino cooling. 
The left-hand panels show the maximum value of $\xi_l$ (top left) and  $\Delta \widetilde{\mu}_l$ (bottom left) throughout the core as a function of time.
The right-hand panels show $\xi_l$ (upper right) and  $\Delta \widetilde{\mu}_l$ (bottom right) as functions of density for a given time that is marked by vertical dotted lines in the left-hand panels.
This time approximately corresponds to the moment when $\xi_l = 5.5$.

The lower panels of Fig.~\ref{Fig:delta_mu} justify our assumption $\Delta \mu_l \ll \mu_n, \mu_p, \mu_l$. 
Indeed, the maximum deviation from beta equilibrium does not exceed 0.05~MeV. 
Such small deviations also explain why the proton fraction at a given density remains very close to its beta equilibrium value (cf. the left panel in Fig.~\ref{Fig:Yp}).
We can also see that $\xi_\mu$ always exceeds $\xi_e$ (the same for  $\Delta \widetilde{\mu}$, obviously).
It is interesting to note that despite a noticeably different thermal evolution for NSs with DD2 and NL3 EOSs (see Fig.~\ref{Fig:Cool_EOS}), their chemical evolution is rather similar. 
$\Delta \widetilde{\mu}$ is larger for the NL3 EOS, but since this star is overall warmer, $\xi_\mathrm{NL3}$ is actually lower than $\xi_\mathrm{DD2}$.

\looseness=-1
Another point worth discussing here is the spatial distribution of $\Delta \widetilde{\mu}$ (the lower right panel).
We have not taken diffusion into account; thus, we cannot explicitly verify the common assumption made in the literature that $\Delta \widetilde{\mu}$ is constant throughout the core.
One can see that without diffusion, $\Delta \widetilde{\mu}$ is not really constant and has a complex dependence on the spatial coordinate (density in our case).
However, implementing diffusion goes far beyond the scope of this paper.
Unlike the diffusion, the assumption of a constant redshifted internal temperature ($\widetilde{T}$) is straightforward for us to check, and we can confirm that, indeed, the internal temperature is virtually constant throughout the whole volume of the star (except for the heat blanketing envelope).
Neither radial nor angular relative variations of $\widetilde{T}$ exceed $10^{-4}$ at $t \geq 1$~Myr, and the largest variations occur in the crust. 
In the core they are even smaller.  

The constancy of $\widetilde{T}$ explains why the spatial distributions of $\xi_l$ and $\Delta \widetilde{\mu}_l$ look very similar.
The only other interesting thing to note here is that these distributions have different monotonicity for the NL3 and DD2 EOSs.

\section{Conclusions} 
\label{sec:Concl}

We have performed pioneering simulations of the rotochemical heating phenomenon in full GR in 2D axially symmetric configurations without relying on the slow rotation approximation.
We have explicitly tracked the thermal, mechanical, and chemical evolution of rotating NSs as they spin down, taking into account the 2D (radial and angular) distributions of temperature, composition, and mechanical structure.
As a first step, we employed the simplest model: only nucleonic composition, only the non-equilibrium modified Urca process [Eqs.~\eqref{Eq:beta_decay} and \eqref{Eq:beta_capture}] as the main heating/cooling mechanism, and no superfluidity. 
We also disregarded diffusion of nucleons caused by deviations from chemical equilibrium.
Employing the simplest model of dipolar spin down, we analyzed the impact of the magnetic field strength as well as the effects of the heat blanketing envelope composition, the initial rotation frequency, the NS mass, and the core EOS.

Our conclusions regarding the impact of the magnetic fields, the heat blanketing envelope composition, and the initial rotation frequency coincide with the results of previous works on this topic. 
However, we have found a strong dependence of the cooling curves (during the rotochemical heating phases) on the core EOS, whereas in the literature this dependence was regarded as very weak.
The most likely explanation for this discrepancy is that the authors of the previous works have not considered EOSs with a wide enough range of variation of the symmetry energy as a function of density.
   
The present work examines the simplest possible configuration of rotochemical heating. 
There are several ways in which we plan to improve it.
First, the effects of superfluidity must be taken into account.
The NSs in question are sufficiently cold for superfluidity to play an important role.
The interplay between superfluidity and magnetic fields in the core is also of interest \citep{Kantor_2021}.
The equilibrium modified Urca emissivities can be upgraded to more modern ones, for example, by employing the nuclear width approximation~\citep{Alford_2024_nwa}. 
Second, more exotic compositions involving hyperons can be considered. 
Weak and especially strong non-leptonic reactions are much faster than the semi-leptonic beta processes \citep{Langer_1969}.
They restore partial chemical equilibrium very quickly, whereas the final relaxation to full chemical equilibrium still occurs via the Urca reactions.
Consequently, one may think that the presence of hyperons does not, in fact,  affect the rotochemical heating.
However, this is not the case. 
As we have demonstrated using various EOSs, the equilibrium $y_p(n_\mathrm{B})$ dependence is very important for the rotochemical heating. 
The presence of hyperons may significantly modify this dependence and, therefore, the rotochemical heating.   
Thus, hyperonic matter requires an explicit investigation.


\begin{acknowledgments}
M.B. and A.R. acknowledge support from a grant from the Ministry of Education and Research, CNCS/CCCDI–UEFISCDI, Project No. PN-IV-P1-PCE-2023-0324 and partial support from Project No. PN 23 21 01 02.
D.P.’s work is supported by a UNAM-DGAPA grant PAPIIT-IN114424.
\end{acknowledgments}

\software{
{\ttfamily NSCool 2D VarRot} (not publicly available),
{\ttfamily python} \citep{python},
{\ttfamily matplotlib} \citep{matplotlib},
{\ttfamily f90wrap} \citep{f90wrap},
{\ttfamily numpy} \citep{numpy},
{\ttfamily scipy} \citep{scipy}
}

\appendix

\section{Notes on numerical implementation}
\label{App:Numeric}

The main solver for the thermal evolution equations is described in Appendix~B of Paper~I.
Here we provide some details regarding the mechanical structure and chemical composition solvers. 

First of all, for a given NS, we calculate in advance the mechanical structure and metric potentials on a dense grid of rotation frequencies using \texttt{LORENE}.
Then, for any particular  frequency, we interpolate in this pre-computed grid.   
As discussed in Sect.~\ref{ssec:Structure}, the expression~\eqref{Eq:EnclosedNBar} is the main equation required to track the evolution of the mechanical structure.
The integral involved is computationally challenging. 
For the core of an NS, the integration is numerically very robust.
However, it becomes increasingly unstable as one moves to the outer layers of the crust.
The main reason is that the crust contains a negligible amount of mass compared to the core, making accurate integration there very difficult.
Furthermore, the integrand is the result of the above-mentioned interpolation of the \texttt{LORENE} data, which may not be sufficiently smooth (see the discussion in Sect.~\ref{ssec:EOS}).
So, we calculate the integral in a hybrid manner.
For a given angle, the radial integration is performed employing the (adaptive) Romberg's method with the relative tolerance set to $10^{-8}$.
Afterwards, the angular integration is carried out using the fact that
\begin{align}
	\int_0^{\pi/2} f(\theta) \sin \theta d\theta = \frac{C_0}{2} + \sum_k \frac{C_k}{1-4 k^2},
	\label{Eq:FFT_Int}
\end{align}
where $\{C_k\}$ are the Fourier coefficients of $f$, which we compute by means of the Fast Fourier Transform (FFT).
Although angular integration is not adaptive, it allowed us to filter out (some of) the numerical noise by disregarding all Fourier coefficients that are smaller than $10^{-8}|C_0|$ and stopping the sum when the coefficients no longer decrease in absolute value (due to numerical noise).
Such a complicated scheme was chosen by trial-and-error  because it provided the most robust numerical results.
For example, we tried to perform the angular integration with Romberg's method, but the results depended on the number of angular grid points. 
We also tried the Gauss-Kronrod quadrature as it is supposed to be superior to Romberg's method, but it was significantly slower, most likely because the \texttt{LORENE} data is not smooth enough. 

After finding a way to reliably compute the integral~\eqref{Eq:EnclosedNBar}, we employed the Newton-Raphson method with the derivatives obtained by finite differences to calculate the new values of  $r^i_\mathrm{max}$ and $r^i_\mathrm{min}$ for a given value of $N_i$.
This constitutes our mechanical structure solver. 

The chemical equilibrium equations~\eqref{Eq:chem_evol} are discretized in time by means of the fully implicit backward Euler scheme (the same as the main solver) and solved by the Newton-Raphson method using finite-difference derivatives.

In all cases, including Eq.~\eqref{Eq:Omega_t}, the timesteps for the evolution in time are determined by the main solver.
We have verified afterwards that the rotation frequency, the mechanical structure, and the chemical composition change no more than a few per cent from one timestep to another.
Temperature variations are controlled by the main solver.

\bibliography{RotChem}
\bibliographystyle{aasjournalv7_link}

\end{document}